\documentclass[prb,showpacs,twocolumn]{revtex4}
\usepackage{graphicx}
\usepackage{color}

\begin{document}

\title{Doubling of the critical temperature of FeSe observed in point contacts }

\author{Yu.G.\ Naidyuk $^{1,2}$}
\author{G.\ Fuchs $^{2}$}
\author{D.A.\ Chareev $^{3,4}$}
\author{A.N.\ Vasiliev $^{4,5,6}$}

\affiliation{$^{1}$ B. Verkin Institute for Low Temperature Physics and Engineering,
National Academy of Sciences of Ukraine, 47 Nauky Ave., 61103, Kharkiv,
Ukraine}
\affiliation{$^{2}$ Leibniz-Institut f\"ur Festk\"orper-
und Werkstoffforschung Dresden e.V.,Postfach 270116, D-01171 Dresden, Germany}

\affiliation{$^{3}$ Institute of Experimental Mineralogy, Russian Academy of Sciences, 142432 Chernogolovka, Moscow District, Russia}

\affiliation{$^{4}$ Institute of Physics and Technology, Ural Federal University, Mira str. 19, 620002 Ekaterinburg, Russia}

\affiliation{$^{5}$ Low Temperature Physics and Superconductivity Department, Physics Faculty, M.V. Lomonosov Moscow State University, 119991 Moscow, Russia}

\affiliation{$^{6}$ National University of Science and Technology "MISiS", Moscow 119049, Russia}

\begin{abstract}
Rise in superconducting critical temperature T$_c$ more than two times (exceeding 20\,K)
is discovered in point-contacts created between iron-chalcogenide FeSe single crystal and Cu.
The possible reasons of such T$_c$ increase in point-contacts are discussed.
The most probable cause for this may be the interfacial carriers doping
and/or interfacial enhanced electron-phonon interaction.
\end{abstract}



\maketitle

\section{Introduction}
The superconducting (SC) compound FeSe, having the simplest crystal structure among other
SC iron chalcogenides and pnictides, has attracted a great interest during the last years.
Primarily, this is connected with the possibility of a large alteration of their SC critical
temperature T$_c$.  Thus, the moderate T$_c$ of about 8\,K in bulk FeSe \cite{Hsu08} increases drastically
up to 27 K under pressure \cite{Mizu08} and achieves even a maximum of 36.7\,K at 8.9\,GPa \cite{Medv09}.
Such T$_c$ enlargement is unlikely to be found for any other SC material. On the other hand,
T$_c$ climbs up to tremendous 109\,K in the case of a FeSe monolayer on SrTiO$_3$, as it has been
shown by means of in situ electrical transport measurements  \cite{Ge15}. After that, various
methods to increase T$_c$ were utilized. One of them is doping of the FeSe topmost layer by
excess electrons by covering the surface using alkali metals \cite{Miyata15}
or applying the liquid-gating technique \cite{Lei15}. However, as it was shown
recently \cite{Seo15} such doping has only a moderate effect, increasing T$_c$ up to 20\,K of FeSe bulk crystal.
That is, substrate-interfacial effects play the main role in increasing T$_c$ in the case
of FeSe monolayer on SrTiO$_3$, likely due to interface enhanced electron-phonon interaction.
Although FeSe monolayer has very high T$_c$, they survive, so far, only in-situ at high
vacuum condition, while using of a protection layer suppresses high-T$_c$ superconductivity
in FeSe  monolayer \cite{Ge15}. Recently, an increase of T$_c$ almost twice than that of the bulk
crystals, was reported for atmosphere-stable FeSe films with a practical thickness of about
several hundred nanometers \cite{Qiu15}. This increase has been explained by proper tuning the
Fe-vacancy disorders via changing the Fe/Se ratio. In this communication,
we report about the observation of more than doubling of T$_c$ onset in point-contacts (PCs)
created on a bulk FeSe single crystal. We believe that our results provide
helpful information in order to understand in more detail the role of the interface
to modify the properties of superconducting FeSe.

\section{Experiments and Results}
The plate-like single crystals (flakes) of FeSe$_{1-x}$ ($x$=0.04$\pm$0.02, \#CD-946) superconductor
were grown in evacuated quartz ampoules  using AlCl$_3$/KCl flux technique in a permanent temperature
gradient as described in Ref.\cite{Char13}. Resistivity and magnetization measurements revealed a SC transition
temperature T$_c$ up to 9.4\,K \cite{Char13} and an onset of superconductivity at about 10\,K.

PCs were established by touching of a sharpened thin Cu wire (${\o}$=0.2\,mm) to the $ab$-plane
of FeSe cleaved by a scalpel at room temperature or contacting by the wire an edge of the plate-like samples.
Thus, we have measured heterocontacts between normal metal Cu and the FeSe crystal mostly along  two directions.

The differential resistance $dV/dI(V)= R(V)$ of PC
was recorded by sweeping the $dc$ current $I$ on which a small $ac$ current $i$ was superimposed using a
standard lock-in technique. The measurements were performed in the temperature range from 3\,K up
to 25\,K. No principal difference in $dV/dI(V)$ data was observed for "plane" or "edge" PC geometry,
because $dV/dI(V)$ differ more significantly from one PC to another.

Typical $dV/dI(V)$
data for PCs with different resistance are shown in Fig.\,1 in our previous publication \cite{Naid16}.
For low-Ohmic PCs with resistance up to several Ohms the main feature in $dV/dI(V)$ is a pronounced
sharp zero-bias minimum \cite{Naid16}. For the overwhelming majority of PCs, independently,
either $dV/dI(V)$ demonstrate additionally occasional Andreev-reflection like features or not
\footnote{ The presence or absence of the characteristic Andreev-reflection features in $dV/dI$ curves allows
to conclude about regime (ballistic, diffusive or thermal)\cite{Naid05} of current flow in superconducting PCs.
In general, the critical temperature has no relation to this regime of current flow through a PC.
Moreover, the heating and critical current effects do not play a role at T$_c^{onset}$ determination, 
since the disappearance of superconductivity in PC we can monitor watching the minimum 
in $dV/dI$ at zero bias (or at zero current). On the other hand, the critical 
temperature itself can be a criteria of the PC quality.
When it is lower than the bulk value, it means that the PC quality is low. }
(see, e.g. Figs. 4 \& 5 in Ref. \cite{Naid16}), this minimum and accompanying side
maxima disappear at temperature around 10\,K \footnote{As a critical temperature onset
T$_c^{onset}$ in PC, we took the temperature at which the zero-bias minimum disappears,
as was marked by the dotted circle in Figs. 1\&2.}. This range was a bit (1--2\,K) higher
for some of the PCs. Unexpectedly, we have found PCs (one PC created by "edge" geometry and one PC
created on $ab$-plane), where SC features were observed to break down only above 20\,K, as it is shown
in Figs.\,1 \& 2. The statistic of the variation of T$_c^{onset}$ with the PC resistance is shown in Fig.\,3.

Thereby, we claim to observe a doubling of the local T$_c^{onset}$ in a restricted geometry of the FeSe crystal,
which is determined by the PC size. The latter was estimated in our previous paper \cite{Naid16} for a PC with similar
resistance to be in the range of 0.1-1$\mu$m. The large uncertainty in the determination of the PC size is
due to the unknown residual resistivity of FeSe at the interface with Cu and due to the heterocontact
geometry, where the partial filling of the PC volume by one (FeSe) or another (Cu) material is unknown.
\begin{figure}[t]
\includegraphics[width=0.5\textwidth]{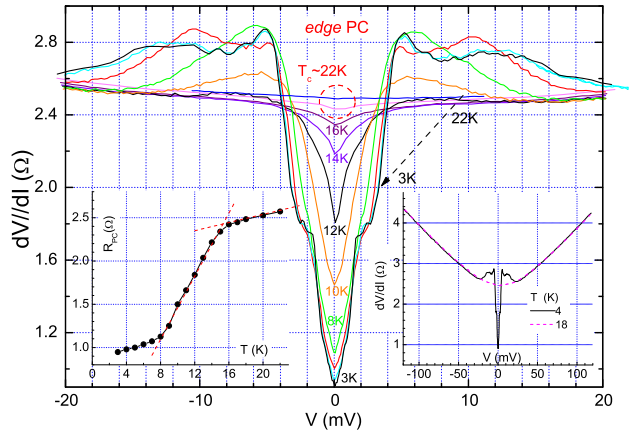}
\caption{(Color online) Evolution of $dV/dI$ curves with temperature for a PC created by touching of an
edge of a FeSe crystal by a Cu wire. The upper two curves marked by the circle are measured at
18 and 22\,K (top curve). Right inset shows two $dV/dI$ curves measured at 4 and 18\,K for a larger bias. Left inset
shows the dependence of the PC resistance at zero-bias versus temperature. Dotted lines are given to guide the eye.
\label{fig:fig1}}
\end{figure}

\section{Discussion}
The first thought that comes to mind about increasing of T$_c$ in PC is that it is connected with a pressure effect.
Indeed, the small size of a PC (or more precisely speaking, the small contact area, which, in general, can be
larger than that of metallic contact itself due to, e.g., surface oxides), which can be in the order of a few microns,
make it possible to cause large pressure (by mechanical creation of PC) within the PC core. According to Fig.\,4 in Ref.\cite{Mizu08},
to reach the onset critical temperature of about 20\,K, the pressure should exceed 1\,GPa.  At the same time, we believe
that metallic Cu wire cannot produce a pressure larger than the yield strength of Cu. The latter reaches only about 0.07\,GPa
\footnote{See "The Engineering Tool Box" web site http://www.engineeringtoolbox.com/young-modulus-d\_417.html}
and cannot be much larger at low temperature.

Another observation that contradicts the pressure explanation of
the T$_c$ increase comes from Fig.\,3. Intuitively it is clear that the pressure in a PC is expected to be larger
for PCs with larger resistance or respectively smaller size. Contrary, as shown in Fig.\,3, two PCs with smaller
resistance (larger size) exhibit a two-times higher T$_c^{onset}$ than the other PCs. Probably, the pressure effect is
responsible mainly for the T$_c^{onset}$ scattering between 10 and 14\,K as seen in Fig.\,3.
\begin{figure}[t]
\includegraphics[width=0.5\textwidth]{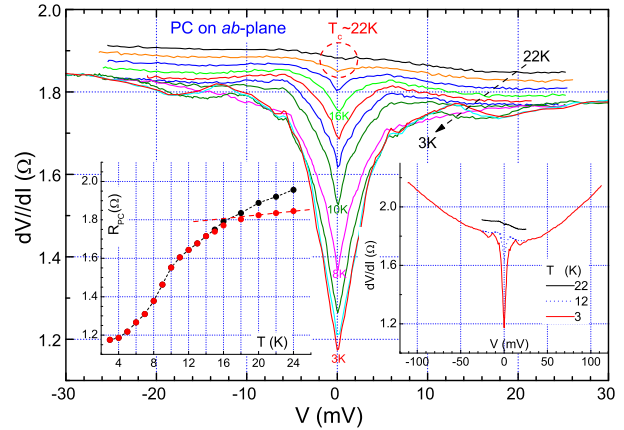}
\caption{(Color online) Evolution of $dV/dI$ curves with temperature for PC created by touching of $ab$-plane of FeSe
flake by a Cu wire. The upper two curves marked by circle are measured at 20 and 22\,K (top curve). Right inset shows $dV/dI$ measured
at 3, 12 and 22\,K for a larger bias. Left inset shows dependence of the PC resistance at zero-bias versus temperature.
Full (red) circles show corrected PC resistance after subtracting from resistance increase above 12\,K
(as it is seen from the shifting of $dV/dI$ "background" at larger bias) coming, likely,
from the bulk.  Dotted line is given to guide the eye.
\label{fig:fig2}}
\end{figure}

An enhancement of T$_c$ in PCs was observed also in Co-doped Ba-122 \cite{Sheet10} and
in FeTe$_{0.55}$Se$_{0.45}$ \cite{Park10} iron-based superconductors.
In both cases this enhancement was about 30\%, that is much less than we observe. In the first case, the authors suppose the formation
of phase-incoherent quasiparticle pairs at a temperature well above T$_c$ arising from strong fluctuation of the phase of the complex
superconducting order parameter. In the second case, the authors assume also novel quasiparticle scattering due to strong antiferromagnetic
spin fluctuations. As to our case, we do not believe that such kind of fluctuations is able to increase T$_c$ on 100\%.

Different scenario to explain the T$_c$ rise in FeSe films may be a fortunate arrangement of iron vacancies somewhere
at the interface, as assumed in Ref.\cite{Qiu15}. However, T$_c^{onset}$ in that FeSe films was found to increase only up to about
15\,K that is 5\,K lower than in our case. Nevertheless, it cannot be excluded also the joint effect of pressure
and iron vacancy arrangement, that might result in 20\,K onset superconductivity in our PCs. However, high
pressure in PC will result rather in disarrangement of the Fe vacancies.

Alternatively,
the observed  T$_c$ increase may be an interfacial effect due to additional doping \cite{Seo15} and/or interface-enhanced
electron-phonon coupling \cite{Ge15} from the side of the normal metal.  Thus, a real understanding of the observed
T$_c$ rise in PCs on the base of FeSe is a challenging task. In this case, the restricted PC geometry and
interfacial effects can play a decisive role.
\begin{figure}[t]
\includegraphics[width=0.5\textwidth]{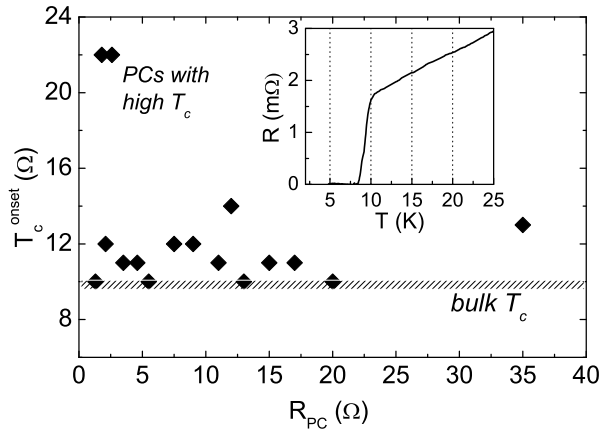}
\caption{Statistic data of T$_c^{onset}$ variation for PCs with different resistance. Horizontal stripe marks
T$_c^{onset}$ for a bulk FeSe taken from the resistance data shown in the inset. Inset: resistance versus
temperature of FeSe sample from Ref.\cite{Char13}.
\label{fig:fig3}}
\end{figure}

\section{Conclusion}
In summary, we have investigated the nonlinear conductivity of PCs on the base of FeSe single crystals due to the transition
into the SC state. We found that SC features in the differential resistance $dV/dI(V)$ persist up to  20\,K for some PCs.
Such doubling of the local critical temperature in FeSe cannot be explained only by pressure effects in the PC.
As a possible explanation a suitable arrangement of iron vacancies is discussed along with the presence of pressure.
Apparently, the underlying physical nature of the observed effect can be understood by taking into account the
restricted geometry of the PC core and interfacial effects.

\section*{Acknowledgements}
Yu.G.N. acknowledges financial support of Alexander von Humboldt Foundation in the frame of a research Group linkage program.
Yu.G.N. would like to thank G. E. Grechnev for the stimulating discussion on iron-chalcogenide superconductors and K. Nenkov
for technical assistance. G.F. acknowledges support of the German Federal Ministry of Education and Research within the project
ERA.Net RUS Plus: No146-MAGNES financed by the EU 7$^{th}$ FP, grant No 609556. A.N.V. acknowledges support of the Ministry of
Education and Science of the Russian Federation in the frames of Increase Competitiveness Program of NUST "MISiS" (No K2-2014-036)
and Russian Foundation for Basic Research (No 14-02-92002). D.A.C. and A.N.V.:  Supported by Act 211 Government of the
Russian Federation, agreement No 02.A03.21.0006. Funding by the National Academy of Sciences of Ukraine under project $\Phi$3-19
is gratefully acknowledged.

\end{document}